\documentclass[aps,prc,twocolumn,superscriptaddress]{revtex4}
\usepackage{graphicx}
\usepackage{dcolumn}
\usepackage{bm}
\begin{document}
\title{Improved radiative corrections for $(e,e'p)$ experiments -- A novel approach to multi-photon bremsstrahlung}
%
\author{Florian Weissbach}
\affiliation{GSI Helmholtzzentrum f\"ur Schwerionenforschung mbH, D -- 64291 Darmstadt, Germany}
\affiliation{Departement f\"ur Physik und Astronomie, Universit\"at Basel, CH -- 4056 Basel, Switzerland}
\author{Kai Hencken}
\affiliation{Departement f\"ur Physik und Astronomie, Universit\"at Basel, CH -- 4056 Basel, Switzerland}
\affiliation{ABB Schweiz AG, Corporate Research, CH -- 5405 Baden-D\"attwil, Switzerland}
\author{Daniela Kiselev}\thanks{n\'ee Rohe}
\affiliation{Departement f\"ur Physik und Astronomie, Universit\"at Basel, CH -- 4056 Basel, Switzerland}
\affiliation{Paul Scherrer Institut, CH -- 5232 Villigen, Switzerland}
\author{Dirk Trautmann}
\affiliation{Departement f\"ur Physik und Astronomie, Universit\"at Basel, CH -- 4056 Basel, Switzerland}
%

%
%
%
\date{\today}
\begin{abstract}
Radiative processes lead to important corrections to $(e,e'p)$
experiments.
While radiative corrections can be calculated exactly in QED 
and to a good accuracy also including hadronic corrections, 
these corrections cannot be included into data analyses to arbitrary orders 
exactly.
Nevertheless consideration of multi-photon brems\-strah\-lung above the
low-energy cut-off is important for many $(e,e'p)$ ex\-pe\-ri\-ments.
To date, higher-order brems\-strah\-lung effects concerning electron 
scattering experiments have been implemented approximately by employing the
soft-photon approximation (SPA).
In this paper we propose a novel approach to multi-photon emission 
which partially removes the SPA from $(e,e'p)$ experiments.
In this combined approach one hard photon is treated exactly; 
and additional (softer) brems\-strah\-lung photons are taken into account
resorting to the soft-photon approximation.
This partial removal of the soft-photon approximation is shown
to be relevant for the missing-energy distribution for several
kinematic settings at {\sc mami} and {\sc tjnaf} energies.
\end{abstract}
%
%
%
\maketitle
\section{Introduction}
\label{intro}
Coincidence electron scattering experiments are an important tool for probing
both nuclear structure and the structure of the nucleons.
$A(e,e'p)(A-1)$ experiments ($A$ being the atomic number) are {\em e.g.}~used 
to study the independent particle shell model and spectral functions and,
in connection with that, occupation numbers and correlations.
And the $(e,e'p)$ reaction has also been employed to study the structure
of, {\em e.g.},~the proton electric form factor in the so called
Rosenbluth technique, apparently differing from polarization transfer 
measurements \cite{arrington}.\\

\begin{figure}[t]
\centering
\includegraphics[width=8cm]{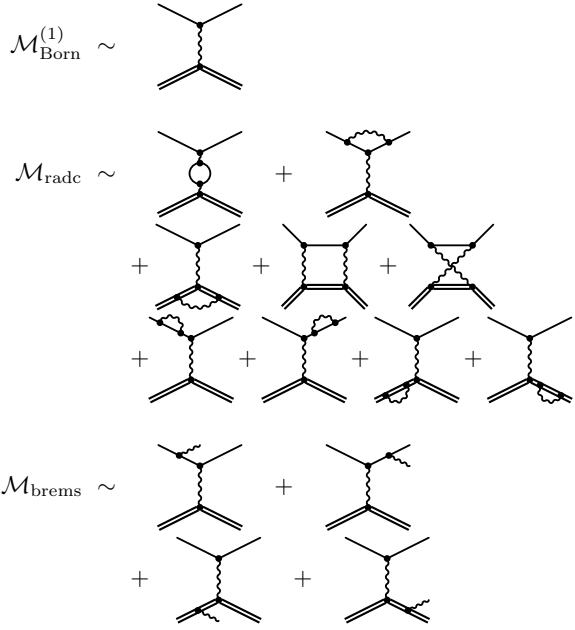}
\caption{\label{fig1} Feynman diagrams beyond the leading order.
The leading Feynman diagram ${\cal M}_{\rm Born}^{(1)}$ together with vertex corrections,
vacuum polarization and bremsstrahlung are called the Born 
approximation.}
\end{figure}

Results from electron scattering experiments are subject to radiative 
corrections, {i.e.}~QED amplitudes going beyond the leading-order Born 
term (see fig.~\ref{fig1}).
These corrections are of relative order $\alpha$, but since they come
with large logarithmic corrections they can contribute significantly to
the cross section.
Vertex correction, vacuum polarization, self-energy diagrams,
and the two-photon exchange (TPE) are referred to as 
{\em internal radiative corrections}.
And the four brems\-strah\-lung diagrams constitute the 
{\em external radiative corrections} and are the main focus of this paper.
By introducing a small parameter associated with the photon energy
resolution of the detectors, $E_0$, one can split up the cross section
into a 'non-radiative part' including vertex corrections,
va\-cuum polarization, self-energy contributions, TPE, and the emission of 
soft brems\-strah\-lung photons with energies below $E_0$;
and into a 'radiative part', accounting for the emission of brems\-strah\-lung
photons with energies above the low-energy cut-off $E_0$
\cite{motsai}.
The individual contributions from the internal and external 
radiative correction diagrams are divergent.
It has first been shown by Schwinger \cite{schwinger}
that by introducing the low-energy cut-off these divergences cancel.\\


The emission of brems\-strah\-lung alters the particle momenta and energies
seen by the detectors and has to be corrected for in data analyses. 
Mo and Tsai first discussed this feature of electron scattering experiments
\cite{motsai,tsai}, considering single-photon brems\-strah\-lung exactly,
aside from an approximation in the calculation of the TPE contribution.
Multi-photon emission is only included for soft photons with energies
smaller than $E_0$.
A comprehensive review on radiative corrections can be found in
ref.~\cite{maximon}.\\

In order to obtain the desired experimental accuracy radiative corrections
cannot be limited to the second order amplitudes.
One has to take into account higher-order brems\-strah\-lung processes
(multi-photon emission) above the low-energy cut-off 
\cite{yennie,gupta,makins} which is straight-forward in soft-photon approximation (SPA).
In SPA, multi-photon emission translates to a simple exponentiation of the
brems\-strah\-lung contribution because of a factorisation of the 
brems\-strah\-lung cross section \cite{yennie,gupta,makins}.
A brems\-strah\-lung photon of energy $\omega^0$ has no backlash on the 
electron-proton scattering in the limit where $\omega^0\rightarrow 0$.
In this limit brems\-strah\-lung amplitudes factorise into the first-order
Born amplitude times the amplitude for emitting a soft photon.
The factorisation also holds for multi-photon emission, if each emitted photon
has vanishing energy.
Summing over all orders of soft-photon emission then leads to exponentiation
which also gives the cross section the correct asymptotic behaviour
as the parameter 
$E_0\rightarrow 0$ \cite{schwinger,maximon,schwinger,nordsieck}.
Strictly speaking the SPA is only valid in the limit where all
brems\-strah\-lung photon energies go to zero.
In data analyses the approximation is applied however to emission of
brems\-strah\-lung photons with finite energies \cite{weissbach}.
The question arises up to which upper bound for the maximum brems\-strah\-lung
photon energy the SPA is a good approximation.\\

While single-photon corrections  to electron scattering
experiments have been calculated exactly,
including ha\-dro\-nic loops to a good accuracy 
\cite{motsai,tsai,makins} 
and including proton structure \cite{maximontjon}, 
multi-photon data analyses are usually performed in SPA. 
It allows both for straight-forward inclusion of 
higher-order brems\-strah\-lung and for straight-forward Mon\-te Car\-lo
generation of the brems\-strah\-lung photon angular distribution
\cite{weissbach}.
The purpose of this paper is to improve the multi-photon
radiative correction treatment for $(e,e'p)$ experiments at
{\sc mami} and {\sc tjnaf} energies by partially removing the SPA from
multi-photon radiative corrections.\\

As the dominant contribution to brems\-strah\-lung in $(e,e'p)$
experiments comes from electron brems\-strah\-lung, we will initially
omit brems\-strah\-lung originating from the proton.
In sect.~\ref{results} we will then include proton brems\-strah\-lung
with recourse to SPA, showing that proton brems\-strah\-lung
is only a minor correction to the full radiative corrections.\\

The two TPE diagrams (see fig.~\ref{fig1}) go beyond the Born approximation and
have received a lot of attention recently in H$(e,e'p)$ experiments
studying the proton electric form factor \cite{arrington,blunden03,guichon03}.
But they are non-relevant for $A(e,e'p)$ nuclear structure experiments where $A>1$.

\section{Multi-photon SPA cross section}
\label{multiSPA}
The emission of a single photon by, {\em e.g.},~the scattered
electron is
\begin{eqnarray}
\label{Mei}
{\cal M_{\rm ef}} &=& 
ie^3{\bar u}(k')\gamma^\alpha \varepsilon_\alpha
\left[ 
\frac{i\gamma^\nu (k'+\omega)_\nu+m}{(k'+\omega)^2-m^2}
\right]
\gamma^\mu u(k)
\nonumber\\
&&\times\frac{1}{q^2}
{\bar v}(p')\Gamma_\mu(q^2) v(p) \, ,
\end{eqnarray}
where $k$ and $k'$ are the four-momenta of the incident and the scattered
electron; and $p$ and $p'$ are the four-momenta of the incident and final proton.
The electron and proton spinors are denoted by $u$ and $v$, respectively.
The momentum transfer to the proton is defined by
\begin{eqnarray}
\label{q}
q\equiv k-k'-\omega \, .
\end{eqnarray}
Here $\omega=\omega^0(1,1,\Omega_\gamma)$ denotes the four-momentum of the
brems\-strah\-lung photon and $\varepsilon$ from eq.~(\ref{Mei}) is its 
helicity.
The vertex
\begin{eqnarray}
\label{Gamma}
\Gamma_\mu(q^2)\equiv
F_1(q^2)\gamma_\mu
+\frac{i\kappa F_2(q^2)}{4M}\sigma_{\mu\nu}q^\nu 
\end{eqnarray}
contains the electromagnetic structure of the target nucleus.
$M$ and $m$ are the proton and electron masses, respectively.
In the limit where the photon energy $\omega^0$ vanishes
the SPA can be applied and the single-photon emission amplitude (\ref{Mei}) 
simplifies yielding \cite{landau,weinberg}
\begin{eqnarray}
\label{Meispa}
{\cal M}_{\rm ei} \sim
e{\cal M}_{\rm Born}^{(1)}
\left(\frac{-k\cdot\varepsilon}{k\cdot\omega}\right)
\, .
\end{eqnarray}
Here the first-order Born amplitude is given by
\begin{eqnarray}
\label{born}
{\cal M}_{\rm Born}^{(1)}
= ie^2{\bar u}(k')\gamma^\mu u(k)
\frac{1}{q^2}
{\bar u}(p')\Gamma_\mu u(p) \, .
\end{eqnarray}
The factorization of the SPA amplitude (\ref{Meispa})
also applies to the other three brems\-strah\-lung diagrams
(see ${\cal M}_{\rm brems}$ in Fig.~\ref{fig1}).
Summing and averaging over the spins and performing the
QED traces, we obtain the factorized single-photon cross section as
\begin{eqnarray}
\label{cross1}
\frac{d^5\sigma}{d\Omega_{\rm e}d\Omega_\gamma d\omega^0}
=\frac{d\sigma^{(1)}}{d\Omega_{\rm e}}\frac{A(\Omega_\gamma)}{\omega^0}\, ,
\end{eqnarray}
where $\frac{d\sigma^{(1)}}{d\Omega_{\rm e}}$ is the elastic first-order 
electron-proton scattering Born cross section and
\begin{eqnarray}
\label{A}
A_{\rm el}(\Omega_\gamma)\equiv
-\frac{\alpha(\omega^0)^2}{4\pi^2}
\left(
\frac{k'}{k'\cdot\omega}
-\frac{k}{k\cdot\omega}
+\frac{p}{p\cdot\omega}
-\frac{p'}{p'\cdot\omega}
\right)^2
\end{eqnarray}
is the SPA angular distributions of the 
brems\-strah\-lung photon \cite{makins}.\\

This distribution (\ref{A}) exhibits peaks as a function of the photon angles 
$\Omega_\gamma$ in the directions of the incident electron, the scattered
electron, and the recoiling proton.
Observing that most of the brems\-strah\-lung is emitted along the 
$e$-direction and the $e'$-direction, Schiff introduced the so-called 
peaking approximation for inclusive $(e,e')$ experiments in 1952 \cite{schiff}
which was extended to exclusive $(e,e'p)$ experiments \cite{makins,simc,mceep},
approximating the angular distribution (\ref{A}) by
\begin{eqnarray}
\label{Apeaking}
A_{\rm el}(\Omega_\gamma)&\sim&
\lambda_{\rm e}\delta(\Omega_\gamma-\Omega_{\rm e})+
\lambda_{\rm e'}\delta(\Omega_\gamma-\Omega_{\rm e'})
\nonumber\\
&&+\,
\lambda_{\rm p'}\delta(\Omega_\gamma-\Omega_{\rm p'}) \, ,
\end{eqnarray}
and can be found {\em e.g.}~in ref.~\cite{makins}.
The functions $\lambda_{\rm e}$, $\lambda_{\rm e'}$, and 
$\lambda_{\rm p'}$ can be obtained by integrating 
the $e$-, $e'$-, and $p'$-contribution to the angular distribution
(\ref{A}).
The peaking approximation can be removed from $(e,e'p)$ data analyses, 
as has been shown in ref.~{\cite{weissbach}}.\\

Now we consider the SPA cross section for multi-photon brems\-strah\-lung
with total photon energy below the low-energy cut-off $E_0$ together with
the emission of $n$ photons with energies above the cut-off, which is given as \cite{makins},
\begin{eqnarray}
\label{cross5}
\frac{d\sigma(n,E_0)}{d\Omega_{\rm e} d\omega_1^0 
d\Omega_{\gamma_1} ... d\omega_n^0 d\Omega_{\gamma_n}}
&=&\frac{d\sigma^{(1)}}{d\Omega_{\rm e}}\exp[-\delta_{\rm soft}(E_0)]
\nonumber\\
&&
\times
(1-\delta_{\rm hard})
\nonumber\\
&&
\times
\frac{1}{n!}
\frac{A_{\rm el}({\Omega_{\gamma_1}})}{\omega_1^0} ...
\frac{A_{\rm el}({\Omega_{\gamma_n}})}{\omega_n^0}
\nonumber\\
&&
\times
\theta(\omega_1^0-E_0)...\theta(\omega_n^0-E_0) \, .
\nonumber\\
&&
\end{eqnarray}
As mentioned above the exponential function accounts for multi-photon emission
below the cut-off energy $E_0$ to all orders. 
$\delta_{\rm soft}(E_0)$ and $\delta_{\rm hard}$ contain the 
brems\-strah\-lung and the internal radiative corrections, respectively.
Their definitions can be found in ref.~\cite{makins}.
Necessary integration techniqes are given, {\em e.g.},~in ref.~\cite{thooft}.\\

In order to obtain an event generator we can sum the cross section (\ref{cross5}) 
over all $n$ photons and we integrate over the photon energies up to an upper 
limit $E_{\rm tot}$, which is a parameter associated with the cuts applied to the
data.
We obtain the photon-integrated cross section \cite{weissbach}
\begin{eqnarray}
\label{cross9}
\frac{d\sigma}{d\Omega_{\rm e}} [\chi_{\rm A}]
&=&\frac{d\sigma^{(1)}}{d\Omega_{\rm e}}
\exp[-\delta_{\rm soft}(E_0)](1-\delta_{\rm hard}) 
\nonumber\\
&&
\times
\sum_{n=0}^{\infty}\frac{1}{n!}\left[\lambda\log
\left(\frac{E_{\rm tot}}{E_0}\right)\right]^n
\nonumber\\
&&
\times
\left[\prod_{i=1}^n\int_{E_0}^{E_{\rm tot}}
\frac{d\omega_i^0 d\Omega_\gamma}{\lambda\omega_i^0\log
\left(\frac{E_{\rm tot}}{E_0}\right)} \chi_{\rm A}^n \right] \, ,
\end{eqnarray}
where $\chi_{\rm A}^n$ is an "acceptance function" of the electron
and photon kinematic variables (for those photons with energies larger than $E_0$), 
$\chi_{\rm A}^n = \chi_{\rm A}^n
(\Omega_{\rm e},\omega_1^0,\Omega_1,\ldots,\omega_n^0,\Omega_n)$,
which is the probability for an event to be seen in 
$\frac{d\sigma}{d\Omega_{\rm e}}[\chi_{\rm A}]$ \cite{weissbach}.
In (\ref{cross9}) $E_{\rm tot}$ is chosen such that all brems\-strah\-lung photons
with non-vanishing $\chi_{\rm A}^n$ are included in the integration,
and $\lambda$ is the integral of $A(\Omega_\gamma)$ over the angular distribution,
see Eg.~(\ref{A}),
\begin{eqnarray}
\label{lambda}
\lambda\equiv\int d\Omega_\gamma A_{\rm el}(\Omega_\gamma) \, .
\end{eqnarray}
Cross section (\ref{cross9}) is expressed in terms of a probability density
function which can be used to generate events with certain photon energies, 
multiplicities, and angular distributions using Monte Carlo event generators 
\cite{weissbach}.
Cross section (\ref{cross9}) is independent of the cut-off $E_0$ in the 
limit where $E_0$ becomes small (see Ref.~\cite{makins}).\\

The Monte Carlo routine we introduce in this paper generates multiple 
photons according to cross section (\ref{cross9}).
The photon multiplicities $n$ follow a 
Poisson distribution and the brems\-strah\-lung energies are essentially
distributed according to $1/\omega^0_i$ \cite{weissbach}.
The brems\-strah\-lung photon angles are generated according to
the elastic angular distribution $A_{\rm el}(\Omega_\gamma^i)$ 
from eq.~(\ref{A}), using the elastic (e,e'p) kinematics for the electron and proton momenta,
by a rejection algorithm.
The index $i$ indicates that each individual photon from a multi-photon
event follows the angular distribution (\ref{A}).\\

The SPA simplifies multi-photon brems\-strah\-lung calculations
considerably by the factorisation of the cross section. 
$(e,e'p)$ data analysis procedures additionally adjust 
the kinematic settings of the $(e,e'p)$ reaction. 
And the form factors are evaluated at the adjusted value of $q^2$
\cite{makins,simc,mceep}, following a suggestion by Borie and Drechsel 
\cite{borie,borie2}.
We refer to this extended version of the SPA as the 'modified SPA', or mSPA,
which accounts for the brems\-strah\-lung photons' backlash
on the electron-proton scattering process approximately via a modification
of the kinematics.\\

\section{Multi-photon bremsstrahlung event generation 
beyond SPA}
\label{beyondSPA}
\begin{figure}[t]
\centering
\includegraphics[width=8cm]{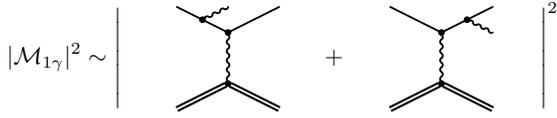}
\caption{\label{fig3} 
The $1\gamma$ matrix element squared.}
\end{figure}

One way of extending the SPA for single-photon brems\-strah\-lung
is to calculate the cross section not in elastic kinematics but in
the real $1\gamma$-ki\-ne\-ma\-tic settings shown in fig.~\ref{fig3}:
Given the photon's four-momentum and the beam energy and 
fixing {\em e.g.}~the electron scattering angles, one calculates the 
remaining kinematic quantities.
The cross section is then calculated using the new $1\gamma$-kinematic 
variables, additionally inserting the modified value of $q^2$ into the 
form factors.\\

Data analyses cannot include an exact multi-photon QED treatment, since the 
exact multi-photon amplitudes cannot be included into the calculations
to arbitrarily high orders.
Therefore we suggest a novel {\em combined approach}: 
we generate multi-photon brems\-strah\-lung according to the SPA distribution.
We choose one hard photon from each scattering event which we treat exactly 
by calculating the exact QED $1\gamma$ matrix element 
(see fig.~\ref{fig3}).
And we treat the remaining photons as soft photons
(with energies still larger than the low-energy cut-off $E_0$), 
employing the SPA.\\

There are several approaches to choosing the hard photon.
We compare four different ones here:
\begin{enumerate}
\item 
The photon with the largest energy $\omega^0$ is chosen as the hard photon.
This method addresses the question of the validity of the SPA.
In this approach the choice for the hard photon depends on the choice of
the reference frame, in contrast to the other approaches described here.
\item 
Calculating the momentum transfer squared, $q^2_{1\gamma}$,
for each photon of the multi-photon event we choose the photon yielding a 
value of $q^2_{1\gamma}$ which deviates most from the elastic value, 
$q^2_{\rm el}$.
In this method form factors are evaluated using this $q^2_{1\gamma}$.
\item 
Calculating the momentum transfer in multi-photon kinematics,
$q^2_{\rm tot}$, the photon is chosen which leads to a value of 
$q^2_{1\gamma}$ which is closest to $q^2_{\rm tot}$.
This method and method 2 focus on both the kinematic aspects of
brems\-strah\-lung emission and on the appropriate calculation of the
form factors.
\item 
Choose the 'hard' photon randomly.
This approach was tested in order to demonstrate the influence of the hard
photons on the result.
\end{enumerate}

We now go beyond the SPA by re-weighting our SPA Monte Carlo generator 
described above, given by the ratio of the exact matrix element and the
SPA matrix element.\\

In order to simplify the type face, we absorb the photon energy dependence 
into the angular distribution $A(\Omega_\gamma^i)$,
\begin{eqnarray}
\label{combinedA}
A(\omega_i)\equiv\frac{A(\Omega_i)}{\omega^0_i} \, .
\end{eqnarray}
In SPA the $1\gamma$-brems\-strah\-lung matrix element squared 
for the hard photon (see fig.~\ref{fig3}) then reads
\begin{eqnarray}
\label{matrixspa}
|{\cal M}_{1\gamma}^{\rm SPA}|^2\approx |{\cal M}_{\rm Born}^{(1)}|^2 
A_{\rm el}(\omega) \, .
\end{eqnarray}
Considering single-photon brems\-strah\-lung first, we
assign to each Monte Carlo brems\-strah\-lung event a weight
\begin{eqnarray}
\label{wex1g}
w_{1\gamma}^{\rm ex}\equiv 
\frac{|{\cal M}_{1\gamma}|^2}{|{\cal M}_{\rm Born}^{(1)}|^2 
A_{\rm el}(\omega)} \, .
\end{eqnarray}
This weight divides out the approximate SPA matrix element and replaces
it by the exact matrix element for single-photon radiation.
In contrast to the 'exact weight' $(\ref{wex1g})$, the mSPA weight
measures the influence of the mSPA,
\begin{eqnarray}
\label{wmod1g}
w_{1\gamma}^{\rm mSPA}\equiv
\frac{|{\cal M}_{\rm el}^{\rm mSPA}|^2 A_{\rm mSPA}(\omega)}
{|{\cal M}_{\rm Born}^{(1)}|^2 A_{\rm el}(\omega)} \, .
\end{eqnarray}
As the exact weight (\ref{wex1g}) this factor re-weights the event but
now for the case of the SPA matrix elements in modified kinematics.
So the first-order Born amplitude ${\cal M}_{\rm el}^{\rm mSPA}$
as well as the photon distribution $A_{\rm mSPA}(\omega)$ in the 
{\em numerator} are evaluated in $1\gamma$-kinematics
(or mSPA kinematics which is the same in this case).\\

This approach can be extended to multi-photon brems\-strah\-lung.
Assuming that the $n$ brems\-strah\-lung photons have been ordered such that
the $n$th photon is the hard photon, we define the exact weight as
\begin{eqnarray}
\label{wexng}
w_{n\gamma}^{\rm comb}\equiv 
\frac{|{\cal M}_{1\gamma}(\omega_{\rm hard})|^2 
A_{\rm mSPA}(\omega_1)...A_{\rm mSPA}(\omega_{n-1})}
{|{\cal M}_{\rm Born^{(1)}}|^2 A_{\rm el}(\omega_{\rm hard})
A_{\rm el}(\omega_1)...A_{\rm el}(\omega_{n-1}) } \, ,
\nonumber\\
\end{eqnarray}
where $A_{\rm mSPA}$ is the angular photon distribution in modified SPA;
and we define the modified weight as
\begin{eqnarray}
\label{wmodng}
w_{n\gamma}^{\rm mSPA}&\equiv&
\frac{|{\cal M}_{\rm Born}^{(1){\rm mSPA}}|^2}
{|{\cal M}_{\rm Born}^{(1)}|}
\nonumber\\
&&\times
\frac{ A_{\rm mod}(\omega_{\rm hard}) 
A_{\rm mod}(\omega_1)...A_{\rm mod}(\omega_{n-1})}
{A_{\rm el}(\omega_{\rm hard})
A_{\rm el}(\omega_1)...A_{\rm el}(\omega_{n-1}) } \, .
\nonumber\\
\end{eqnarray}

The Monte Carlo routine introduced here generates multi-photon events in
SPA.
Each scattering event is assigned two weights (\ref{wexng}) and (\ref{wmodng})
using the four methods for selecting the hard photon, described at the 
beginning of this section.
In order to check the combined approach against data (where available) and
against SPA simulations, we embedded the combined Monte Carlo routine into 
{\sc simc}, a Monte Carlo simulation for electron scattering 
\begin{figure}[t]
\centering
\includegraphics[width=7cm]{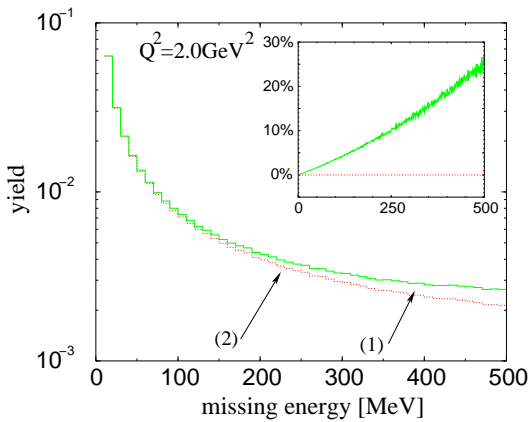}
\caption{\label{fig4} 
Missing-energy distribution for multi-photon bremsstrahlung.
The solid curve (1) (green) was obtained using the modified SPA weight 
(\ref{wmodng}). 
The dotted curve (2) (red) represents the combined approach using
weight (\ref{wexng}).
The inset graph shows the deviation between the two curves in
percent normalized to the combined result.
At $E_{\rm m}=100\,{\rm MeV}$ the modified SPA calculation
overestimates the radiative tail by 3.6\%, 
at $E_{\rm m}=500\,{\rm MeV}$ the deviation is 25\%.
The momentum transfer is $Q^2=0.6\,{\rm GeV}^2$.}
\end{figure}
\begin{figure}[t]
\centering
\includegraphics[width=7cm]{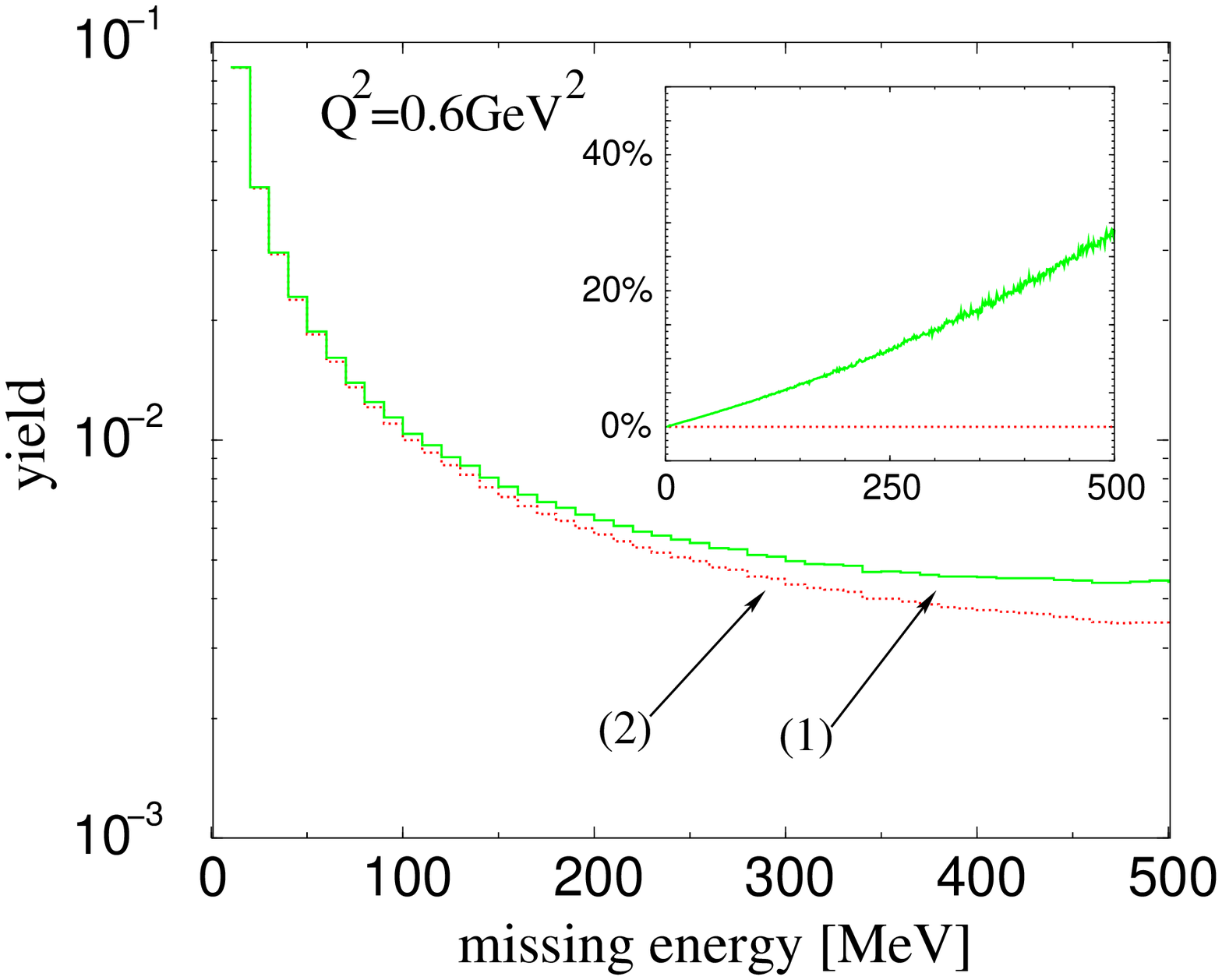}
\caption{\label{fig5} 
Missing energy distribution for multi-photon bremsstrahlung.
The attribution of the curves is as above in fig.~\ref{fig4}.
At $E_{\rm m}=100\,{\rm MeV}$ the modified SPA calculation
overestimates the radiative tail by 3.9\%, 
at $E_{\rm m}=500\,{\rm MeV}$ the deviation is 29\%.
The momentum transfer is $Q^2=2.0\,{\rm GeV}^2$.}
\end{figure}
\begin{figure}[t]
\centering
\includegraphics[width=7cm]{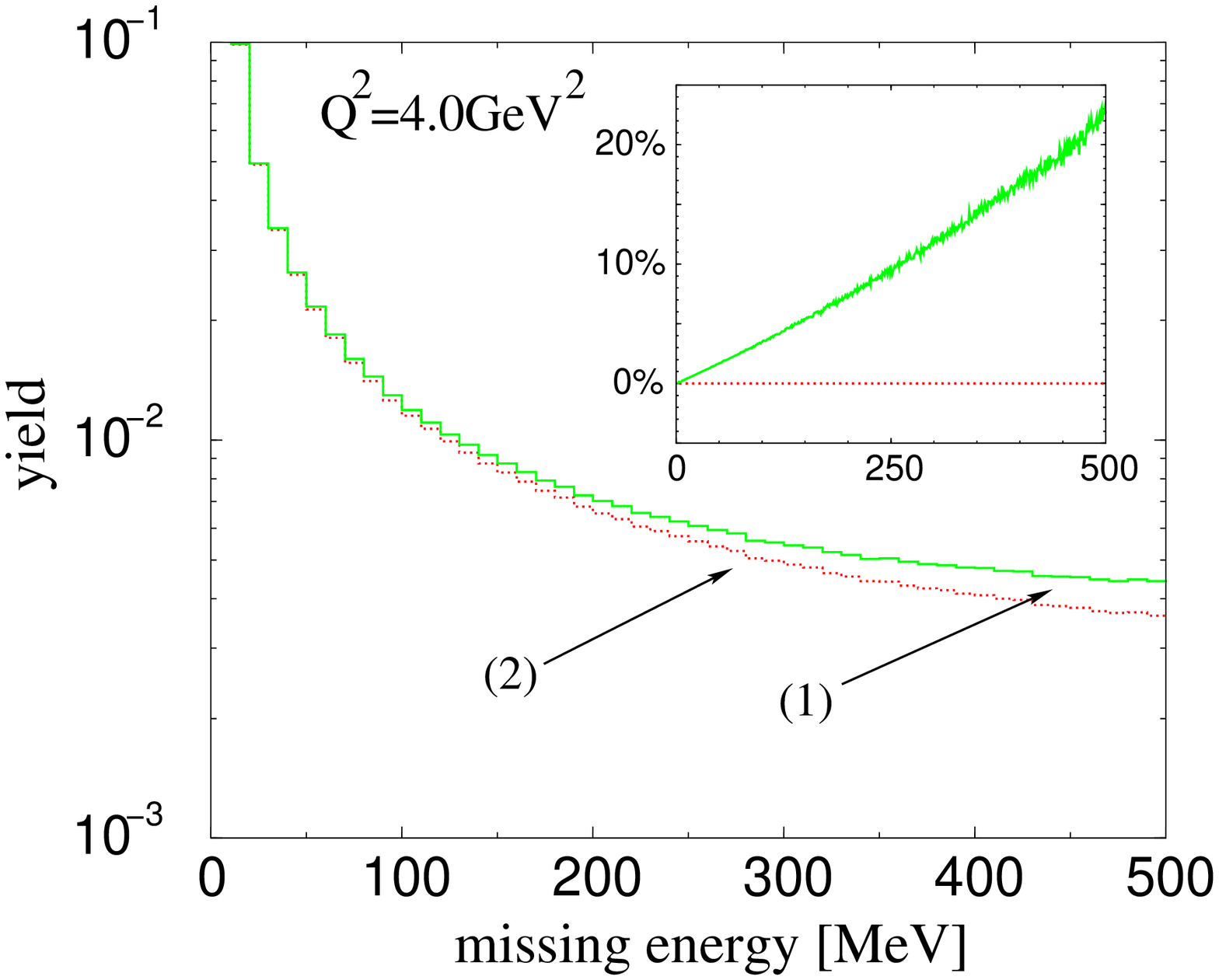}
\caption{\label{fig6} 
Missing-energy distribution for multi-photon bremsstrahlung.
The attributions of the curves is as above in fig.~\ref{fig4}.
At $E_{\rm m}=100\,{\rm MeV}$ the modified SPA calculation
overestimates the radiative tail by 3.5\%, 
at $E_{\rm m}=500\,{\rm MeV}$ the deviation is 23\%.
The momentum transfer is $Q^2=4.0\,{\rm GeV}^2$.}
\end{figure}
\begin{figure}[t]
\centering
\includegraphics[width=7cm]{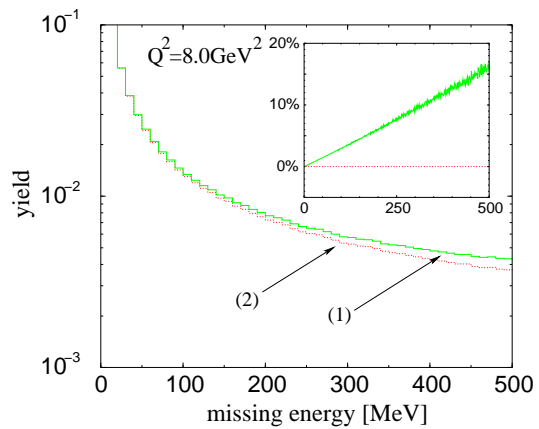}
\caption{\label{fig7} 
Missing-energy distribution for multi-photon bremsstrahlung.
The attribution of the curves is as above in fig.~\ref{fig4}.
At $E_{\rm m}=100\,{\rm MeV}$ the modified SPA calculation
overestimates the radiative tail by 3.0\%, 
at $E_{\rm m}=500\,{\rm MeV}$ the deviation is 17\%.
The momentum transfer is $Q^2=8.0\,{\rm GeV}^2$.}
\end{figure}
ex\-pe\-ri\-ments from {\sc tjnaf} \cite{simc}.
This data analysis code uses a version of the mSPA which -- in contrast to 
our mSPA calculations shown in figs.~\ref{fig4} to \ref{fig7} -- is not 
able to choose a hard photon from a given multi-photon 
brems\-strah\-lung event.
On top of that it makes use of the peaking approximation \cite{schiff,simc}.
Thus we were able to compare the error associated with the use of the SPA 
with other sources of errors.

\section{Results}
\label{results}
To test the combined calculation we simulated the missing-energy distributions
and the photon angular distributions at several kinematic settings.
Ten million events per run were generated in order to get good statistics
for the different yields.
The results turned out to be indistinguishable under approaches 1, 2, and 3
for choosing the hard photon (see previous section).
Only the random choice of the hard photon (approach 4) deviated
from the other methods.
The results presented in this section are therefore based on approach 1.\\

Figures.~\ref{fig4} to \ref{fig7} show multi-photon missing-energy 
distributions, once calculated in mSPA (\ref{wmodng}),
once calculated using the combined approach (\ref{wexng}).
We see that the mSPA calculations overestimate the radiative tails.
While the deviations between the two calculations are of the order
of a few percent for missing energies below $E_{\rm m}=100\,{\rm MeV}$,
the deviations become considerably larger towards the far ends of the
radiative tails.\\

Since we want to evaluate the importance of the SPA in multi-photon
radiative corrections it is important to compare the effect of the improved 
radiative corrections to other sources of errors, present in $(e,e'p)$ 
experiments.
On top of radiative corrections, $(e,e'p)$ experiments are 
{\em e.g.}~corrected for finite detector resolution, for particle decays,
and for multiple scattering \cite{simc,mceep}.\\

\begin{figure}[t]
\centering
\includegraphics[width=7cm]{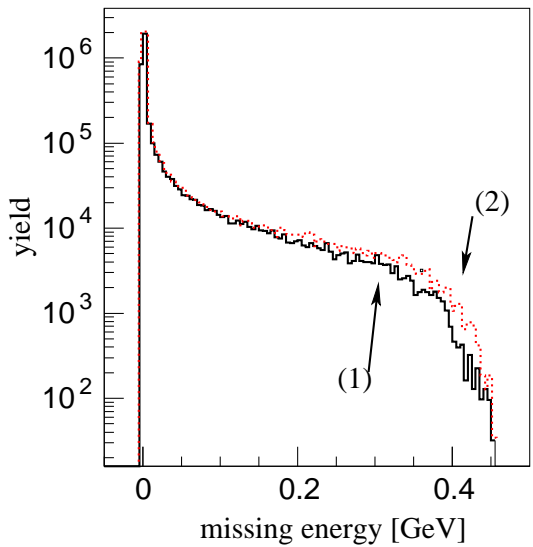}
\caption{\label{fig8} 
Missing-energy distribution for multi-photon bremsstrahlung
simulated with {\sc simc}.
The solid curve (1) (black) represents the standard {\sc simc} modified
SPA radiative corrections.
The dotted curve (2) (red) shows the $E_{\rm m}$ distribution obtained
by inserting our combined radiative correction approach
into {\sc simc}.
The latter one has more strength in the radiative tail.
The total yield differs by 4.6\%.
The momentum transfer is $Q^2=0.6\,{\rm GeV}^2$,
as in fig.~\ref{fig4}.}
\end{figure}
\begin{figure}[t]
\centering
\includegraphics[width=7cm]{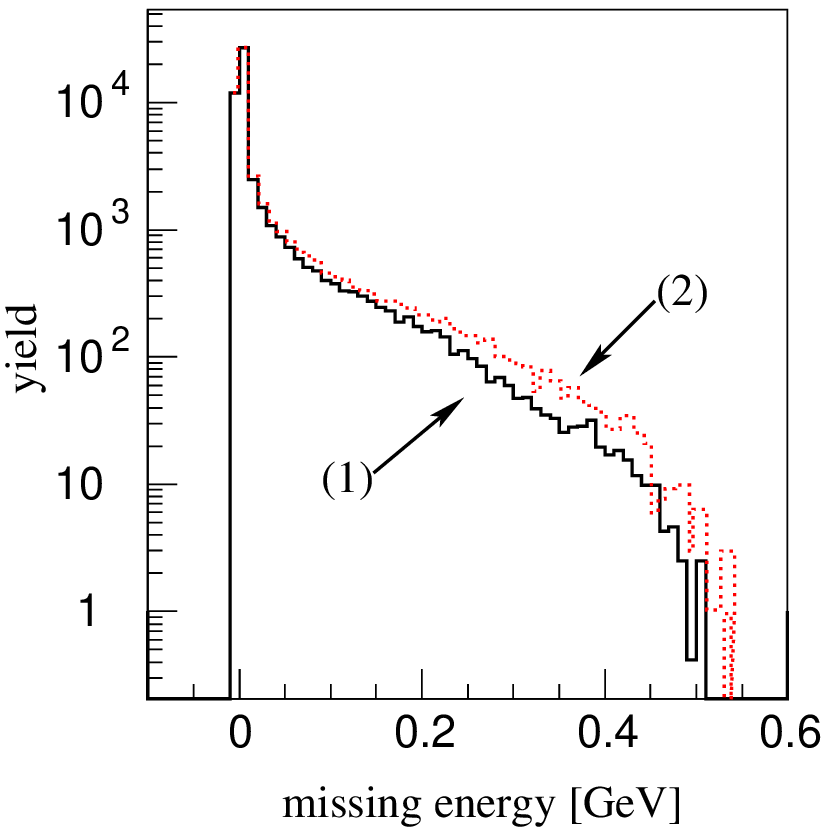}
\caption{\label{fig9} 
Missing-energy distribution for multi-photon bremsstrahlung
simulated with {\sc simc}.
The attribution of the curves is the same as in fig.~\ref{fig8}.
The total yield differs by 2.3\%.
The momentum transfer is $Q^2=2.0\,{\rm GeV}^2$,
as in fig.~\ref{fig5}.}
\end{figure}
\begin{figure}[t]
\centering
\includegraphics[width=7cm]{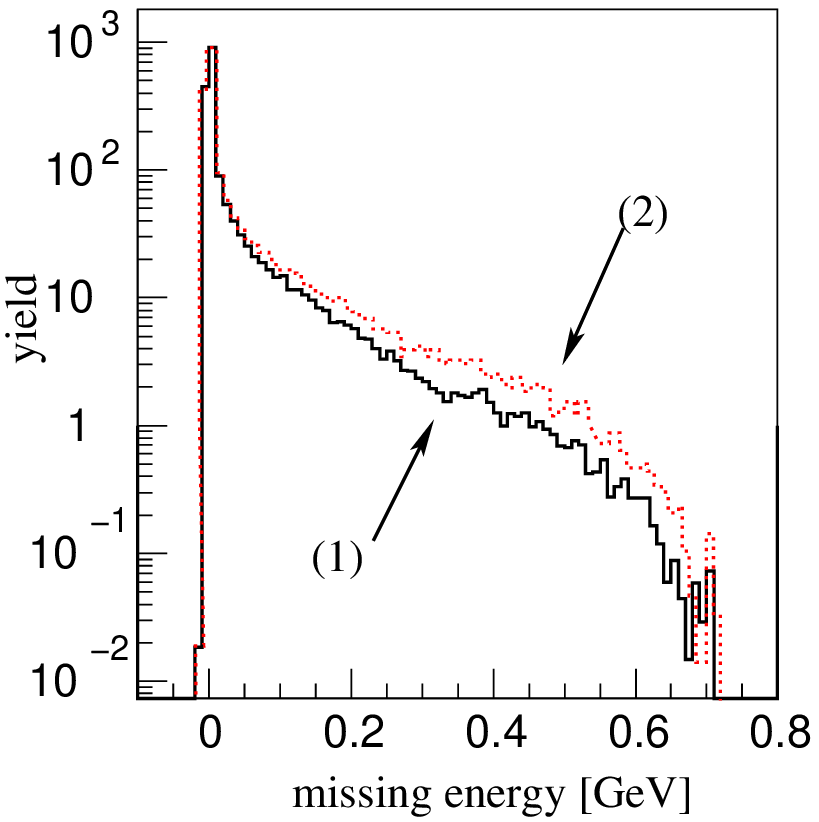}
\caption{\label{fig10} 
Missing-energy distribution for multi-photon bremsstrahlung
simulated with {\sc simc}.
The attribution of the curves is the same as in fig.~\ref{fig8}.
The total yield differs by 1.4\%.
The momentum transfer is $Q^2=4.0\,{\rm GeV}^2$,
as in fig.~\ref{fig6}.}
\end{figure}
\begin{figure}[t]
\centering
\includegraphics[width=7cm]{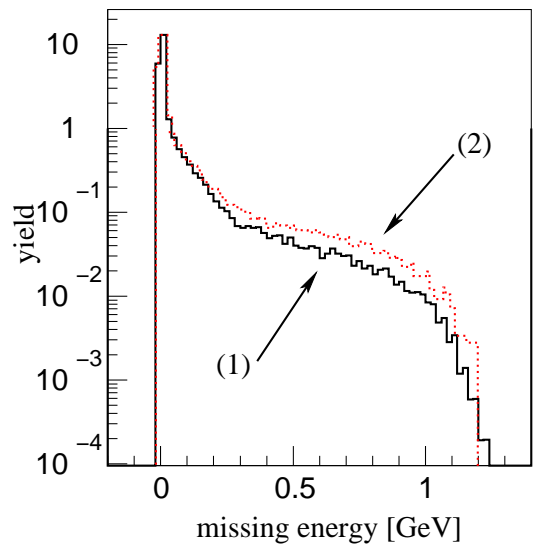}
\caption{\label{fig11} 
Missing-energy distribution for multi-photon bremsstrahlung
simulated with {\sc simc}.
The attribution of the curves is the same as in fig.~\ref{fig8}.
The total yield differs by 1.4\%.
The momentum transfer is $Q^2=8.0\,{\rm GeV}^2$,
as in fig.~\ref{fig7}.}
\end{figure}

The simulations shown in figs.~\ref{fig8} through \ref{fig11} take into account
radiative corrections, detector resolution, particle decays, and multiple
scattering.
They indicate that the combined approach has an impact on the 
missing-energy distribution which is not diluted by the other corrections.
The radiative tails calculated following the combined approach are stronger 
than the radiative tails obtained with the standard {\sc simc}
radiative correction procedure for all kinematic settings considered.\\

As a second observable we considered the angular distribution of the
brems\-strah\-lung photons.
As for the missing-energy distributions we first looked at radiative
corrections only.
Then we additionally  took into account other corrections.
The photon angular distributions are shown in
\begin{figure}[t]
\centering
\includegraphics[width=7cm]{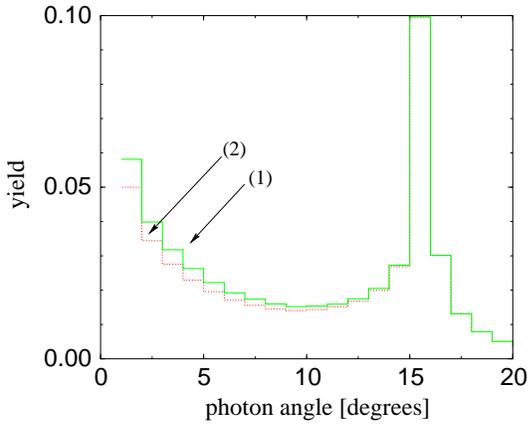}
\caption{\label{fig12} 
Angular bremsstrahlung photon distribution.
The solid curve (1) (green) represents the modified SPA,
the dotted curve (2) (red) shows the combined approach.
The SPA distribution deviates from the combined calculation
especially in the vicinity of the incident electron,
around $0^o$.
The kinematic settings are as in fig.~\ref{fig4}.}
\end{figure}
\begin{figure}[t]
\centering
\includegraphics[width=7cm]{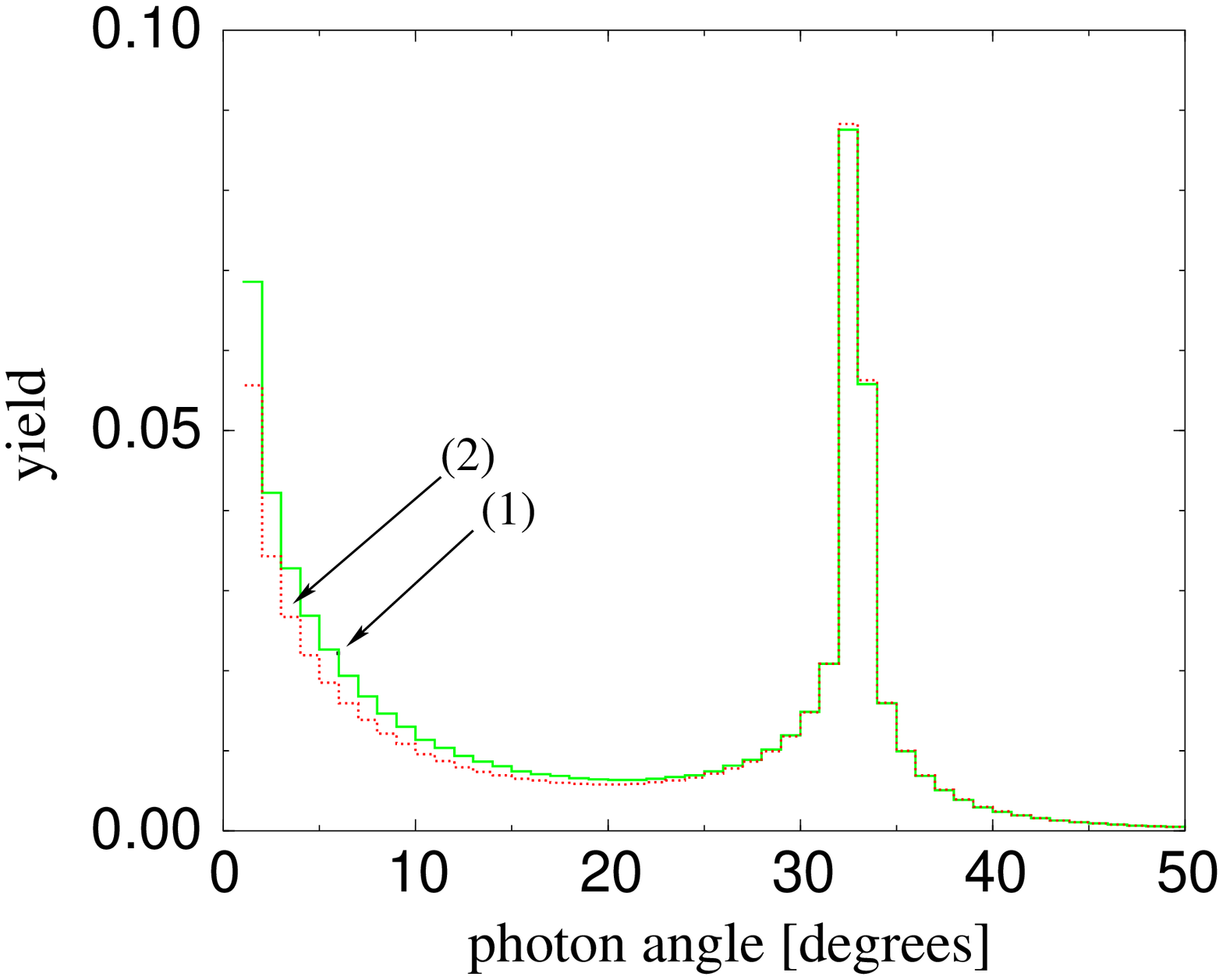}
\caption{\label{fig13} 
Angular bremsstrahlung photon distribution.
The attribution of the curves is as in fig.~\ref{fig12}.
The SPA distribution deviates from the combined calculation
especially in the vicinity of the incident electron,
around $0^o$.
The kinematic settings are as in fig.~\ref{fig5}.}
\end{figure}
\begin{figure}[t]
\centering
\includegraphics[width=7cm]{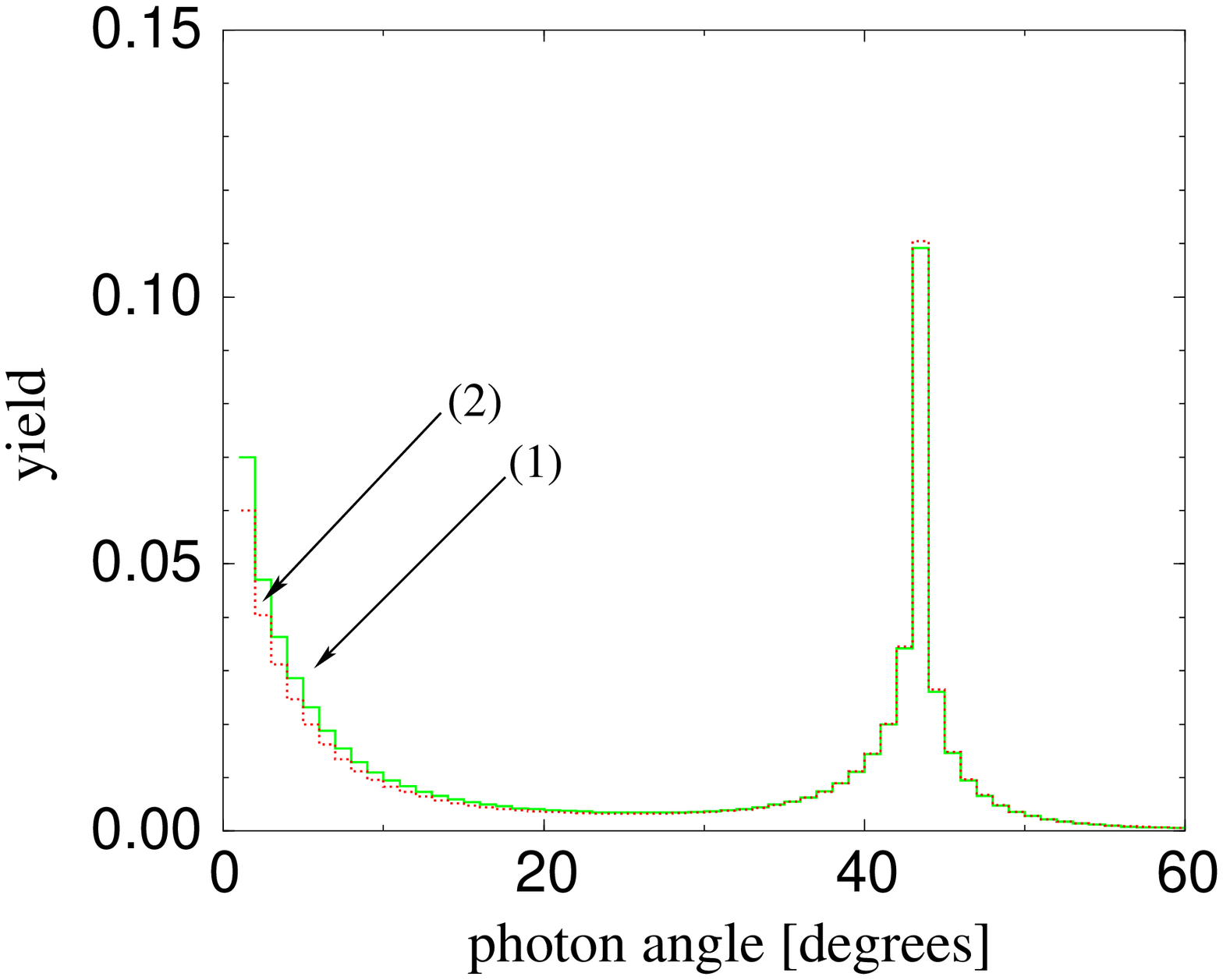}
\caption{\label{fig14} 
Angular bremsstrahlung photon distribution.
The attribution of the curves is as in fig.~\ref{fig12}.
The SPA distribution deviates from the combined calculation
especially in the vicinity of the incident electron,
around $0^o$.
The kinematic settings are as in fig.~\ref{fig6}.}
\end{figure}
\begin{figure}[t]
\centering
\includegraphics[width=7cm]{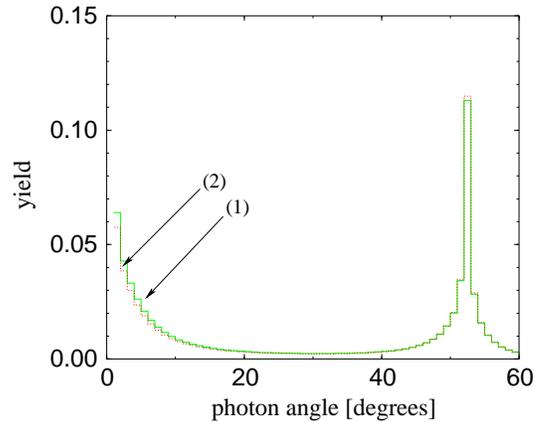}
\caption{\label{fig15} 
Angular bremsstrahlung photon distribution.
The attribution of the curves is as in fig.~\ref{fig12}.
The SPA distribution deviates from the combined calculation
especially in the vicinity of the incident electron,
around $0^o$.
The kinematic settings are as in fig.~\ref{fig7}.}
\end{figure}
figs.~\ref{fig12} to \ref{fig15}.
For all kinematic settings the mSPA calculations overestimate the angular 
distribution in the vicinity of the incident electron, that is small angles.
As for the missing energy the question is
whether this deviation can be seen in the data,
or whether other sources of errors dominate the photon
angular distribution.
In order to determine the impact of the combined approach on
the photon angular distribution we resort to {\sc simc},
as in the case of the missing-energy distribution.\\


Using standard {\sc simc} for comparison implies that the peaking approximation is employed
\cite{simc}.
It is clear that the photon angular distribution, generated in {\sc simc}'s 
standard radiative correction approach 
differs from the combined approach even more than the mSPA 
from the combined calculation shown in figs.~\ref{fig12} to \ref{fig15}
(see also ref.~\cite{weissbach}).
The peaking approximation and thus the standard {\sc simc} analysis
code is known not to describe the experimental angular
distribution accurately in between the two radiation
peaks coming from the incident and the scattered electron
\cite{weissbach}.
\begin{figure}[t]
\centering
\includegraphics[width=7cm]{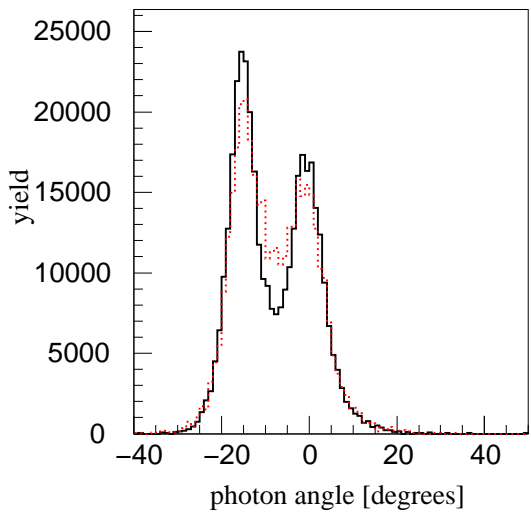}
\caption{\label{fig16} 
Angular bremsstrahlung photon distribution for H$(e,e'p)$ scattering 
generated with {\sc simc}.
The solid curve (1) (black) represents the standard {\sc simc} modified
SPA calculation.
The dotted curve (2) (red) shows the angular distribution simulated with
the combined approach.
The largest difference between the standard {\sc simc} treatment
and the data occurs in the middle between the $e$ and the $e'$
directions and in the height of the two peaks.
This is due to the peaking approximation which is used by the standard
{\sc simc} code; it overestimates the peaks.
The kinematics are the same as in figs.~\ref{fig4} and \ref{fig12}.}
\end{figure}
\begin{figure}[t]
\centering
\includegraphics[width=7cm]{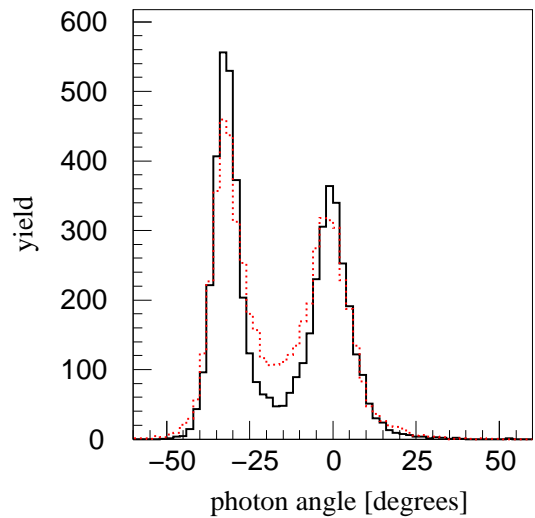}
\caption{\label{fig17} 
Angular bremsstrahlung photon distribution generated with {\sc simc}.
The attribution of the curves is the same as in fig.~\ref{fig16}.
The comparison with experimental data can be found in fig.~\ref{fig20}.
The kinematics are the same as in figs.~\ref{fig5} and \ref{fig13}.}
\end{figure}
\begin{figure}[t]
\centering
\includegraphics[width=7cm]{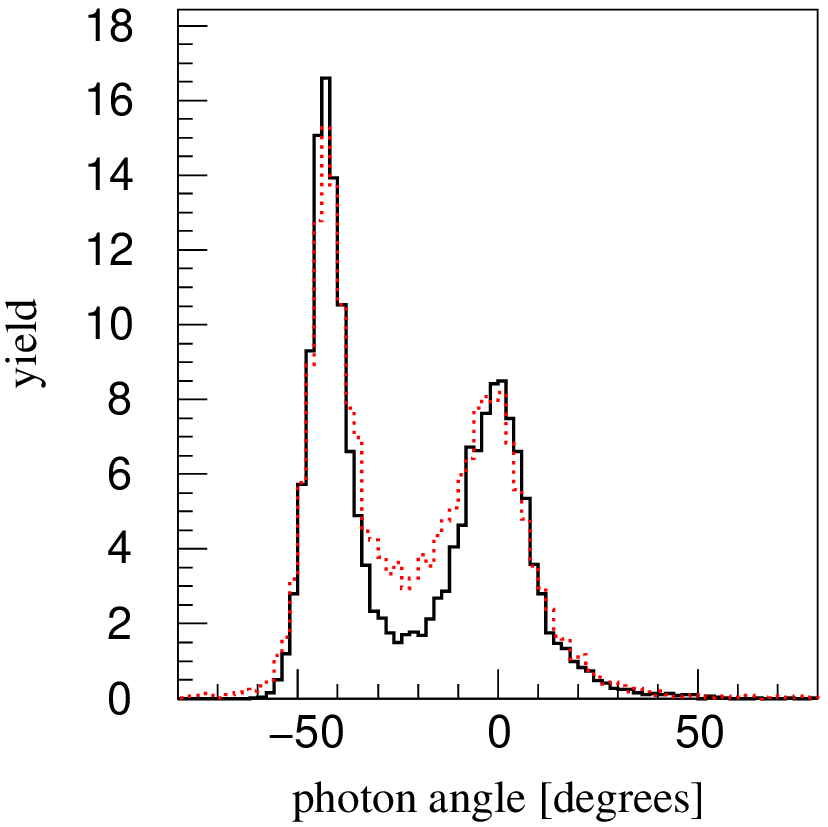}
\caption{\label{fig18} 
Angular bremsstrahlung photon distribution generated with {\sc simc}.
The attribution of the curves is the same as in fig.~\ref{fig16}
but there was no data available for these kinematic settings.
They are the same as in figs.~\ref{fig6} and \ref{fig14}.}
\end{figure}
\begin{figure}[t]
\centering
\includegraphics[width=7cm]{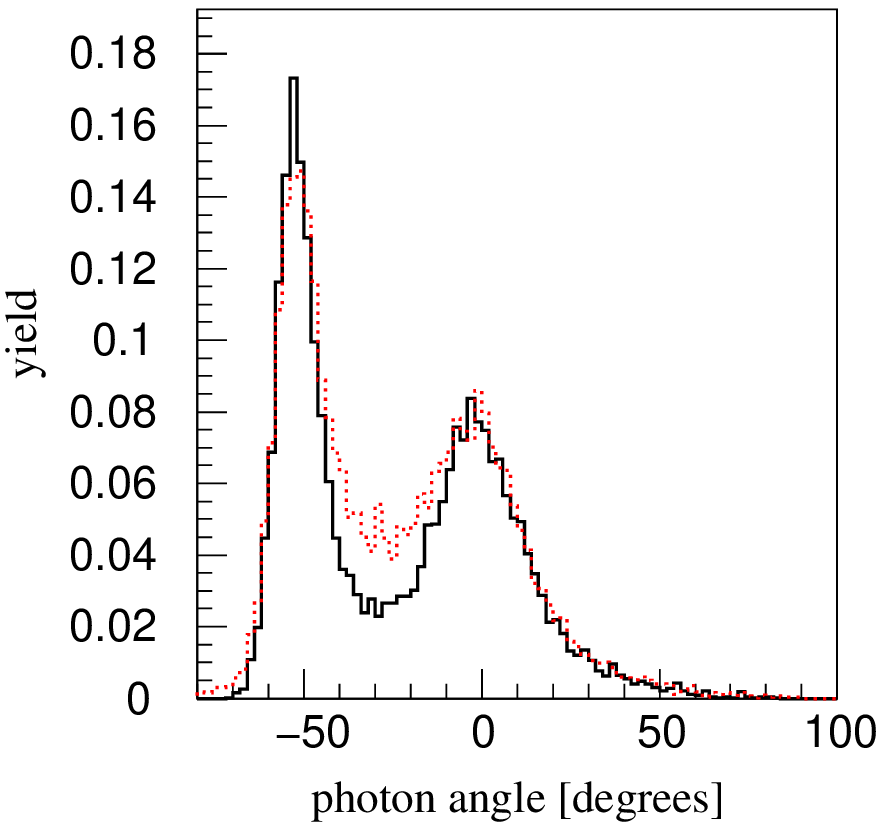}
\caption{\label{fig19}
Angular bremsstrahlung photon distribution generated with {\sc simc}.
The attribution of the curves is the same as in fig.~\ref{fig16}
but there was no data available for these kinematic settings.
They are the same as in figs.~\ref{fig7} and \ref{fig15}.}
\end{figure}
Nevertheless, looking at the photon angular distributions
both with the standard {\sc simc} code and with the modified
version of {\sc simc}, containing our combined simulation,
gives us the opportunity to rate the impact of our approach
compared to the standard {\sc simc} radiative correction procedure.
Where available, we compared the different approaches to data.\\

Figs.~\ref{fig16} to \ref{fig19} show the photon angular distributions
as generated by {\sc simc}. 
The largest deviations between the standard {\sc simc} radiative
correction and the combined approach appear in the middle between the
peaks due to $e$ and $e'$ brems\-strah\-lung.
This is in contrast to figs.~\ref{fig12} to \ref{fig15} where deviations 
occur at small angles.
This indicates that the small deviations between SPA and combined approach 
seen at small angles (see figs.~\ref{fig12} to \ref{fig15}) are washed out 
by other corrections.
The photon angular distribution is not effected by the SPA and by the 
combined approach.
The dominant approximation is the peaking approximation here.\\

\begin{figure}[t]
\centering
\includegraphics[width=7cm]{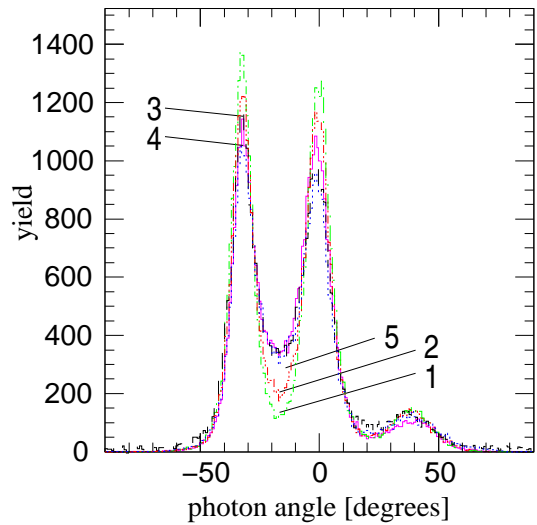}
\caption{\label{fig20} 
Angular bremsstrahlung photon distribution for H$(e,e'p)$ scattering
generated with {\sc simc} and compared to data.
The bump on the right hand side at the proton direction 
is an artefact due to 'punch through effects'.
The dash-dotted curve (1) (green) 
and the (2) (red) curve show standard {\sc simc}
photon distributions, both using different versions of the peaking 
approximation.
The dotted curve (3) (black) represents the data.
The solid line (4) (magenta) is the new combined approach implemented into
{\sc simc}.
Additionally there is a blue line (5) from ref.~\cite{weissbach} showing 
a full angular SPA simulation.
The kinematics are the same as in figs.~\ref{fig5} and \ref{fig13}.
The combined approach (4) fits the data well.}
\end{figure}

Fig.~\ref{fig20} shows experimental data for the photon angular distribution
reconstructed from the H$(e,e'p)$ kinematics shown in figs.~\ref{fig4},
\ref{fig12}, and \ref{fig16} ({\sc tjnaf} experiment E97-006, 
\cite{danielaprl,daniela}) and verifies this conclusion.
This measurement at $Q^2=2.0\,{\rm GeV}^2$ shows that both the mSPA calculation
and the combined approach describe the data well.

In order to additionally include the effect of proton brems\-strah\-lung
in the combined approach, we inserted it, assuming the mSPA is a good 
approximation for proton brems\-strah\-lung, the overall effect of
proton brems\-strah\-lung being small anyway.
The first-order electron brems\-strah\-lung is still calculated exactly.
This was achieved by modifying the weight (\ref{wexng}) into
\begin{eqnarray}
\label{wexngproton}
w_{n\gamma}^{\rm comb}&\equiv&
[|{\cal M}_{1\gamma}(\omega_{\rm hard})|^2 
+ |{\cal M}_{\rm Born}^{(1)}|^2 A_{\rm rest}(\omega_{\rm hard})]
\nonumber\\
&&\times
\frac{A_{\rm mod}(\omega_1)...A_{\rm mod}(\omega_{n-1})}
{|{\cal M}_{\rm Born}^{(1)}|^2 A_{\rm el}(\omega_{\rm hard})
A_{\rm el}(\omega_1)...A_{\rm el}(\omega_{n-1}) } \, ,
\nonumber\\
&&
\end{eqnarray}
where $A_{\rm rest}(\omega_{\rm hard})$ is the angular distribution
(\ref{A}) without the electron-electron terms. 
(In addition (\ref{A}) has been divided by $\omega^0_{\rm hard}$).
We generated the photon angular distribution using this approach, 
comparing it to the combined calculation without the proton (\ref{wexng}).
\begin{figure}[t]
\centering
\includegraphics[width=7cm]{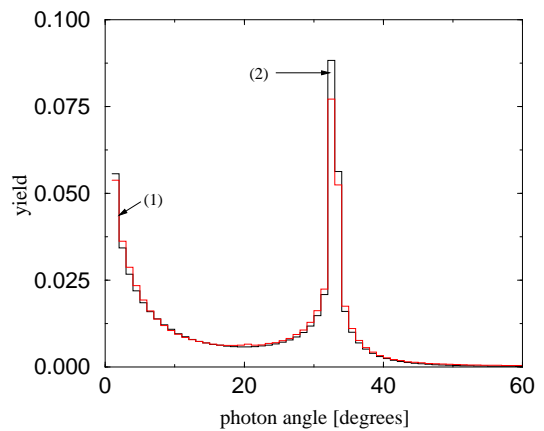}
\caption{\label{figproton} 
Photon angular distribution with (1) (red curve) and without (2) (black curve)
SPA proton brems\-strah\-lung.
The simulation neglecting proton brems\-strah\-lung overestimates
the electron peaks slightly.
In the rest of the photon angle domain the two curves coincide.}
\end{figure}
\begin{figure}[t]
\centering
\includegraphics[width=7cm]{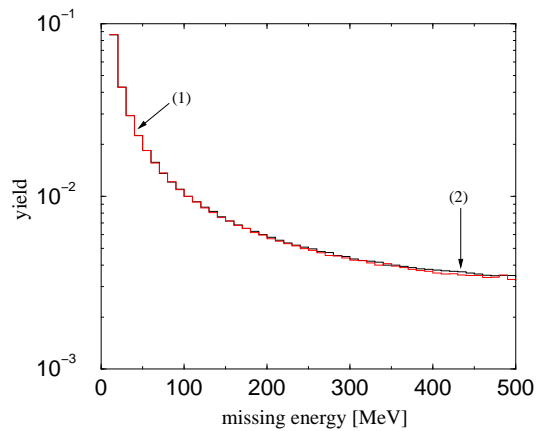}
\caption{\label{figprotonem} 
Missing energy with (1) (red curve) and without (2) (black curve)
SPA proton brems\-strah\-lung.
The two curves are hardly distinguishable.}
\end{figure}
As can be seen in fig.~\ref{figproton}, including mSPA proton brems\-strah\-lung
does not change the photon angular distribution considerably.
Just the electron radiation peaks are slightly overestimated when 
neglecting the proton brems\-strah\-lung.
The missing energy is not changed at all by including proton
brems\-strah\-lung (see fig.~\ref{figprotonem}).
We conclude that the proton brems\-strah\-lung is not relevant
and that mSPA is a good approximation for proton brems\-strah\-lung.\\

We observe that in the case considered here the combined approach reproduces 
the experimental photon angular distribution much better than the standard 
{\sc simc} simulation using the peaking approximation.
The combined approach presented in this manuscript
and the full angular approach from ref.~\cite{weissbach} are much closer to 
each other and to the data than the standard {\sc simc} simulations,
indicating, again, that removal of the peaking approximation is more
important for the photon angular distribution than the partial removal of 
the SPA.

\section{Discussion and conclusion}
\label{discussion}
We have shown that it is feasible to improve $(e,e'p)$ radiative corrections
by partially removing the SPA from multi-photon brems\-strah\-lung 
processes.
Our results are invariant under three different approches to
choosing the hard photon whereas a random choice of the hard photon is 
not feasible.
Treating one photon as a hard photon improves the kinematic treatment of the
$(e,e'p)$ reaction.\\

We compare the combined approach to multi-photon mSPA calculations,
looking at the missing-energy distribution.
The mSPA overestimates the radiative tail for different kinematic settings.
In order to check whether the combined treatment 
has an influence when considering additional experimental
corrections we subsequently inserted our combined calculation into an existing
data analysis code, which in its standard version is using the peaking
approximation and a version of mSPA,
in addition to other corrections \cite{makins,simc}.
We showed that that our combined approach had an impact on
the missing-energy distribution and that it was not hidden
by other sources of background and other effects like, {\em e.g.}~detector resolution.
The computational expense of the combined method was small,
at most a factor of 2, compared to a standard data analysis
code.\\

Similarly, we showed that the photon angular distribution
was overestimated by mSPA especially at small angles,
in the vicinity of the incident electron.
Inserting our combined approach into {\sc simc} we saw large
deviations between the standard {\sc simc} photon angular distribution
and our combined approach.
The bulk of this difference was due to the peaking approximation,
as has already been suggested in ref.~\cite{weissbach}.
The photon angular distribution is hardly sensitive to deviations
originating from mSPA.\\

Our combined approach takes radiative corrections for $(e,e'p)$ experiments 
beyond both the peaking approximation and the soft-photon approximation.
And it treats both the kinematic impact of multi-photon brems\-strah\-lung 
and the evaluation of the form factors at modified momentum transfers more 
systematically than previous $(e,e'p)$ radiative correction procedures.
We have also shown that the SPA is a good approximation for
proton brems\-strah\-lung.


\end{document}